\documentclass[useAMS,usenatbib]{mn2e}
\usepackage[dvips]{graphicx,color}
\usepackage{amssymb}
\usepackage{indentfirst}
\usepackage{epsfig}
\usepackage{pdflscape}
\usepackage{afterpage}
\usepackage{multicol}

\newcommand{\Ka}{\ensuremath{\hbox{K}\alpha~}}

\def\cha{{\it Chandra }}

\def\nus{{\it NuSTAR }}

\begin{document}
\title[]{On the spin-up events and spin direction of the X-ray pulsar GX 301-2}

\author[J. Liu et al.]{Jiren Liu$^{1}$\thanks{E-mail: jirenliu@nao.cas.cn}\\
	 $^{1}$National Astronomical Observatories, 20A Datun Road, Beijing 100012, China\\
 }

\date{}

\maketitle

\begin{abstract}
Recently a retrograde neutron star 
is proposed for the classical wind-fed X-ray pulsar, GX 301-2, to explain the orbital 
spin-up to spin-down reversal near periastron, based on the stream model invoked 
to explain the pre-periastron flare of GX 301-2 previously. We study in detail three 
rare spin-up events detected by Fermi/GBM and find that the spin derivatives are correlated
with the Swift/BAT fluxes, following a relation of $\dot{\nu}\propto F^{0.75\pm0.05}$. 
All the spin-up events of GX 301-2 started about 10 days after the periastron, 
which is the time needed for tidally stripped gas to reach the neutron star. 
The slow rotation of the optical companion implies that
the accreted matter is likely to have angular momentum in the direction of the orbital motion, as in
a Roche-Lobe-like overflow. As a result, the spin-up events of GX 301-2 would favor 
accretion of a prograde disk to a prograde neutron star.
We also find that the flare of intrinsic X-ray emission of
GX 301-2 happened 0.4 days before periastron, while the flare of low energy emission (2-10 keV)
happened about 1.4 days before periastron. The preceding low energy flare
can be explained by stronger absorption of the intrinsic X-ray emission closer to the periastron.
This finding weakened the need of the stream model.
The pulse fraction of GX 301-2 near periastron is reduced heavily, which is likely caused 
by Compton scattering process. Compton reflection from the optical companion might be 
responsible for the observed orbital spin reversal of GX 301-2.
\end{abstract}


\section{Introduction}

In accretion-powered X-ray pulsars, the spinning neutron star with a strong
magnetic field accretes material from the optical companion star and produces 
pulsating X-ray emission. The mass transfer process could be through either 
a stellar wind or a Roche-Lobe overflow or a combination of both.
The transport of angular momentum to the neutron star leads to variations of 
the spin frequency measurable on days timescale,
and therefore, the spin history of X-ray pulsar can be used to probe the mass transfer process
\citep[e.g.][]{RJ77,GL79,Wang81}. 
Some X-ray pulsars show spin-up/spin-down reversals 
separated by intervals from days to decades, a puzzling phenomena not well understood
\citep[e.g.][]{Mak88, Cha97a, Bil97, Mal20}.

Recently, \citet{Mon20} proposed a retrograde neutron star for the classical wind-fed X-ray pulsar,
GX 301-2, which has a long pulse period $\sim680$ s \citep{Whi76}.
While retrograde neutron stars have been suggested to be reminiscent of
the natal kick during the supernova explosion of the progenitor \citep{Hil83,BP95},
the spin direction of neutron star is generally hard to constrain observationally.
The GX 301-2 system has a binary period $\sim41.5$ days and an orbital
eccentricity $\sim0.46$ \citep{Sat86,Koh97}. 
Its optical companion, Wray 977, has a mass within $39-53 M_\odot$ and 
a radius of $62 R_\odot$,
close to the Roche lobe radius near periastron \citep{Kap06}.
The X-ray emission of GX 301-2 exhibits periodic flares about 1.4 days before the periastron 
passage \citep[e.g.][]{Sat86}, and various models, such as equatorial disk and tidal stream, have been 
proposed to explain the pre-periastron flare \citep{PG01,LK08}.
\citet{Mon20} discovered that the spin frequencies of GX 301-2 show an orbital spin-up to 
spin-down reversal near periastron.
If this orbital spin reversal is caused by passage of a tidal stream over the neutron star, 
as modelled by \citet{LK08}, the neutron star should rotate in the opposite direction 
of the orbital revolution \citep{Mon20}.

\begin{figure*}
	\hspace{-1.8cm}
	\includegraphics[width=7.5in]{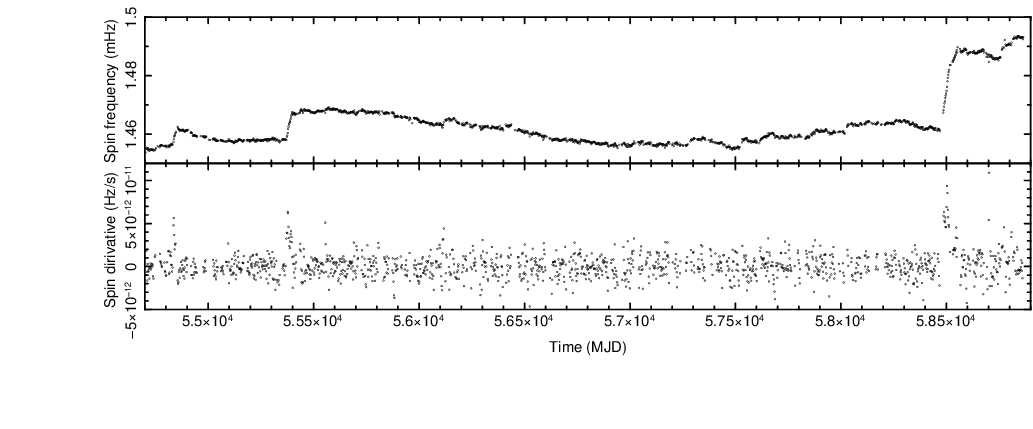}
	\hspace{-0.0cm}
	\includegraphics[width=3.4in]{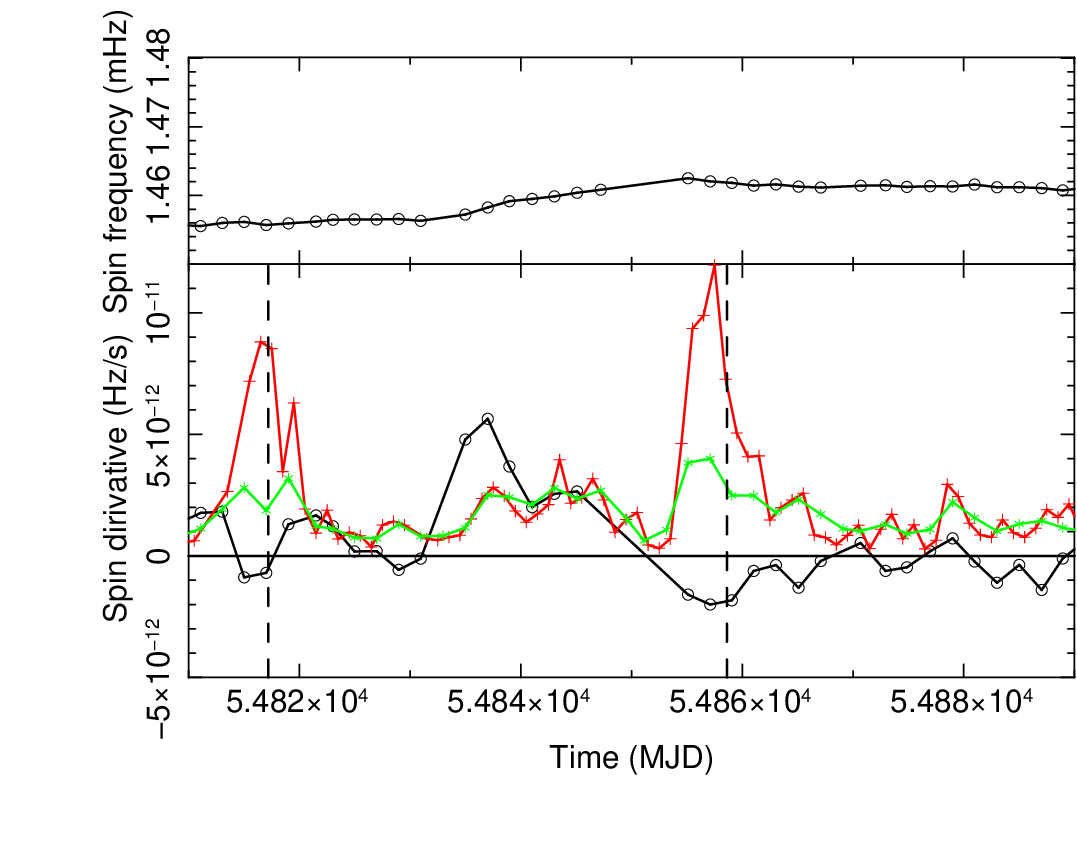}
	\hspace{-0.0cm}
	\includegraphics[width=3.4in]{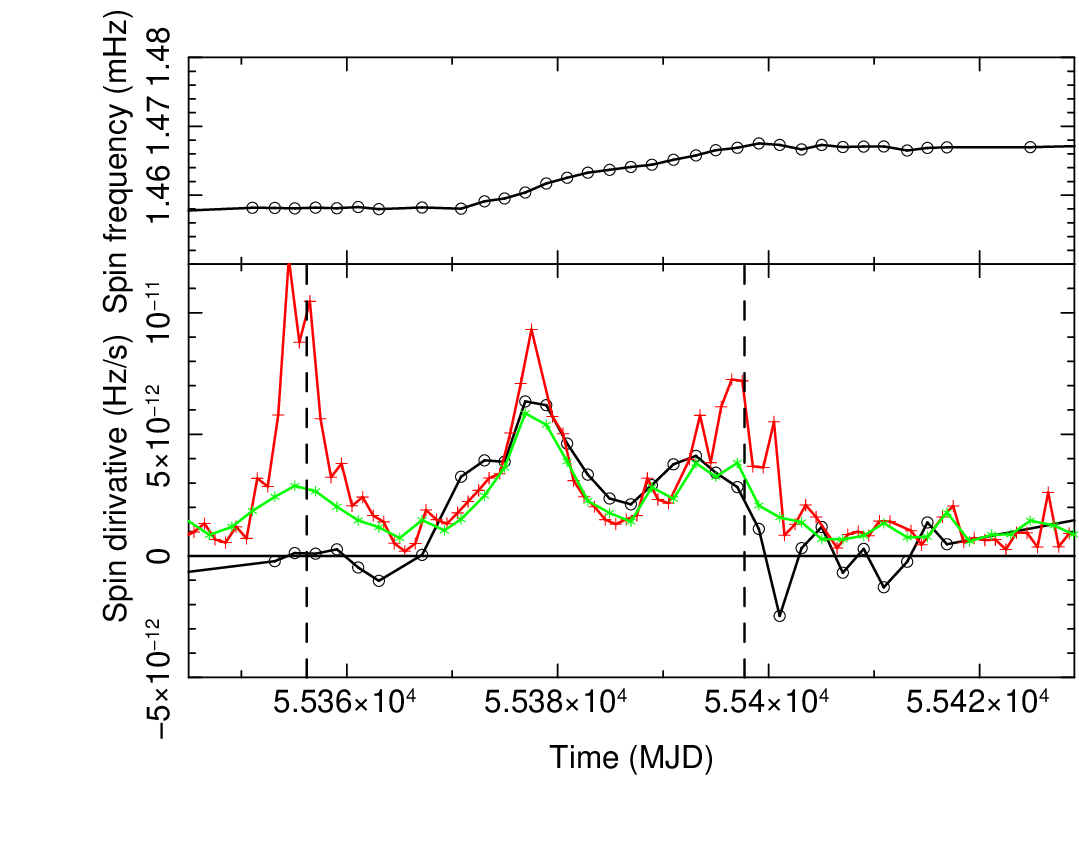}
	\hspace{-0.0cm}
	\includegraphics[width=3.8in]{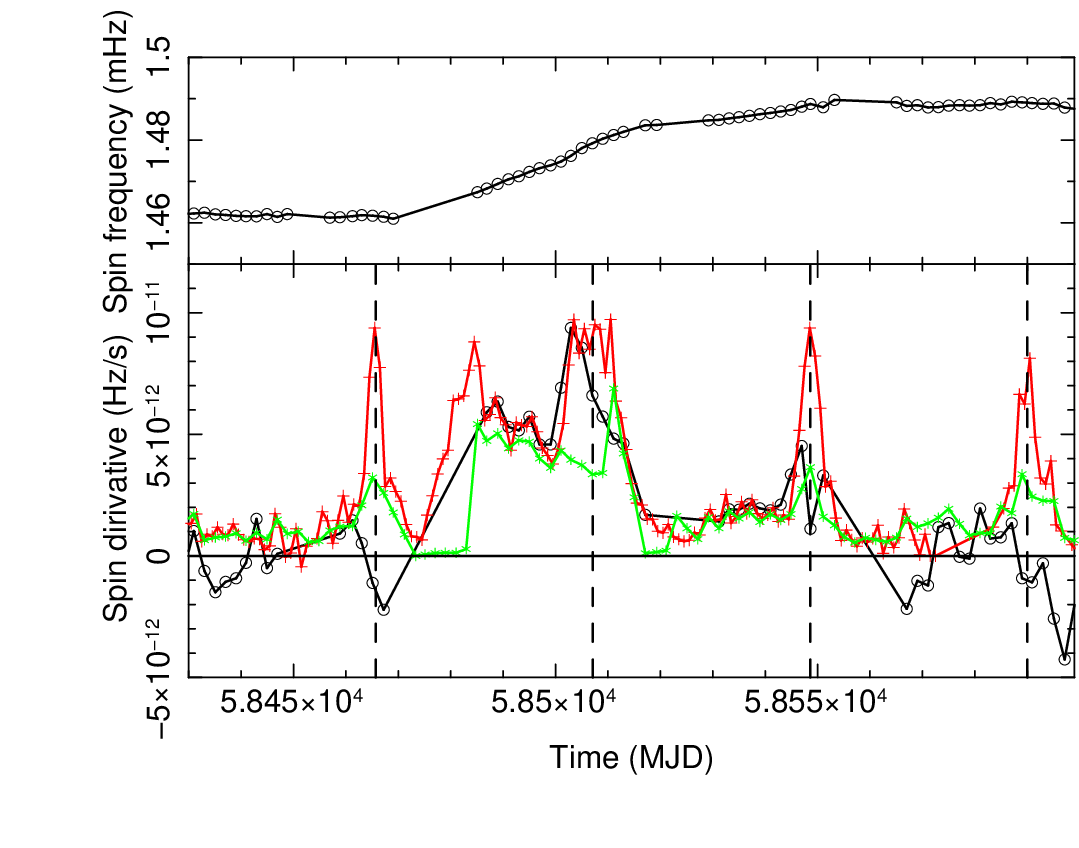}
	\caption{Top: the spin history of GX 301-2 monitored by Fermi/GBM 
since 2008-8-13. Three rapid spin-up events are prominent around MJD  54850, 55400, and 58500.
The corresponding spin derivatives are over-plotted. 
Middle and bottom: zoomed plot of the three spin-up events, together 
with the Swift/BAT 15-50 keV fluxes (red pluses) and the pulsed fluxes within 12-50 keV estimated 
by Fermi/GBM (green stars). Vertical dashed lines indicate the time of periastron.
}
\end{figure*}

Because a case of a retrograde neutron star will have profound implications for the understanding
of X-ray pulsar, in this paper, we study the rare spin-up events of GX 301-2, which 
also provide insights on the spin direction of GX 301-2.
The pulse frequency history of GX 301-2 monitored by the Burst and Transient Source 
Experiment (BATSE) on the Compton gamma ray observatory revealed two 
rapid spin-up events, each lasting for about 30 days, besides erratic variations \citep{Koh97}.
The Gamma-ray Burst Monitor \citep[GBM,][]{GBM09} on the Fermi spacecraft detected 
other three rapid spin-up events, with 
the most powerful one occurred in Jan., 2019. We will study in detail the three rapid spin-up events
detected by Fermi/GBM, together with the data from Swift/BAT and MAXI. 
The last spin-up event has been reported by \citet{Nab19} and \citet{Aba20}.
We will also show
that the flare of intrinsic X-ray emission (15-50 keV) of
GX 301-2 happened 0.4 days before periastron, different from the flare of low energy emission (2-10 keV),
which happened about 1.4 days before periastron.

\section{Spin-up events of GX 301-2}

The spin histories of 
X-ray pulsars are continuously monitored currently with 
Fermi/GBM\footnote{https://gammaray.msfc.nasa.gov/gbm/science/pulsars/}
\citep{GPM09, Mal20}. 
The spin frequencies ($\nu$) of GX 301-2 
monitored by Fermi/GBM are presented in Figure 1, together with the spin derivative, 
$\dot{\nu}$, calculated with a running three points derivative.
As can be seen, the spin frequencies vary on a timescale of days, with three
rapid spin-up events around MJD 54850, 55400, and 58500.
The spin derivatives vary around 0, with an average amplitude 
around $1\times10^{-12}$ Hz\,s$^{-1}$.
The three rapid spin-up events are zoomed in the middle and bottom panel of Figure 1.
The corresponding 15-50 keV fluxes measured by Swift/BAT and 12-50 keV pulsed fluxes 
estimated with Fermi/GBM are also presented in the zoomed plots.
The Swift/BAT data are taken from Swift hard X-ray transient 
program\footnote{https://swift.gsfc.nasa.gov/results/transients/}.
As Fermi/GBM searches the pulsar periodic signal using Fourier method due to the strong 
background, only the pulsed flux is obtained.

\begin{figure}
	\hspace{-0.5cm}
	\includegraphics[width=3.45in]{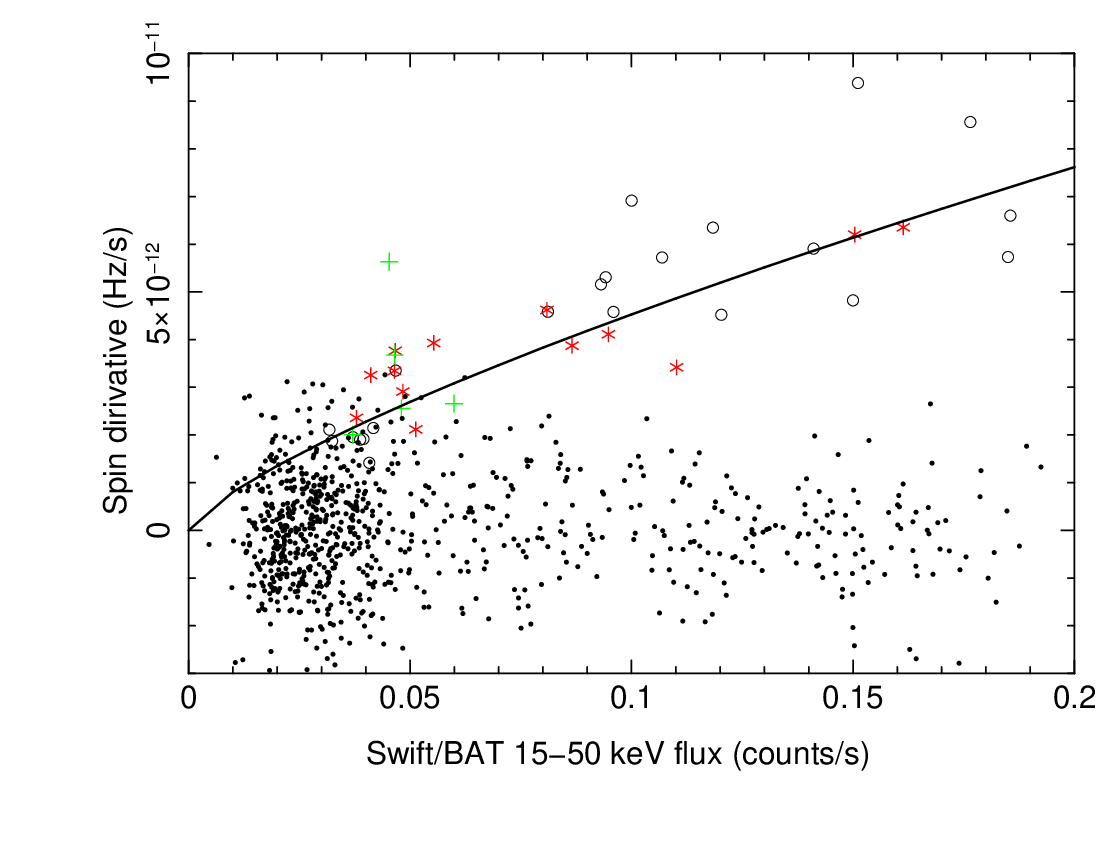}
	\caption{The relation of spin derivatives ($\dot{\nu}$) vs 
		Swift/BAT 15-50 keV fluxes for the first (green pluses), 
		the second (red stars), and the last (open circles) spin-up events. 
		The relation during the no spin-up period (MJD 55400-58400) are over-plotted as black dots.
		The spin derivatives are correlated with the fluxes for the spin-up events, with a fitted 
		relation of $\dot{\nu}\propto F^{0.75\pm0.05}$ (solid line).
				}
\end{figure}

The first spin-up event around MJD 54850 lasts for about 20 days, the second 
event around 55400 lasts for about 30 days, while the last one around 58500 lasts 
for about 80 days. 
As can be seen from the zoomed plot, except for the first 10 days of the first 
event and the periastron period of the last event, the spin derivatives during the 
spin-up periods are well correlated with the Swift/BAT fluxes. To further illustrate this point,
in figure 2, we plot the relation of spin derivative vs Swift/BAT flux for the three spin-up periods,
and for the no spin-up period between the second and the last spin-up events (MJD 55400-58400).
For this purpose, the Swift/BAT fluxes are obtained as the mean value of the orbital 
light curves during the times within which the spin derivatives are calculated.
We see that the spin derivatives within the no spin-up period
show no correlation with the fluxes, while those during the three spin-up episodes stand out
differently and show a positive correlation with the fluxes.
On the other hand, different from a well correlation of pure disk accretion
\citep[e.g., A0535+262 and GRO J1744-28,][]{Bil97}, the data of the spin-up events of 
GX 301-2 are more scattered, which could be due to contamination of wind accretion. 
We fit a power-law to the spin derivatives during spin-up episodes and obtain a relation of 
$\dot{\nu}\propto F^{0.75\pm0.05}$ for 90\% confidence level assuming that the measured 
derivatives have 10 percent errors.

At the beginning of the last spin-up event, the pulsed fluxes are much lower 
than normal values for about $10$ days, and the spin frequencies 
are not measurable by Fermi/GBM. Indeed, as shown by \citet{Aba20}, during this period, 
the pulse fraction within 15-35 keV measured with X-Calibur is only $\sim30\%$.
They noted that the shape of the X-Calibur 15-35 keV pulse profile is 
dominated by one peak, 
quiet different from the normal two-peak profiles observed by \nus during other periods,
and these features can be caused by changes of the accretion and emission geometries.
The column density during this low-pulse-fraction period measured by \citet{Aba20} 
are around $8\times10^{23}$\,cm$^{-2}$, higher than the average value measured around 
these phases \citep[$\sim10^{23}$\,cm$^{-2}$, e.g.][]{IP14}.
It indicates that significant material is present around the neutron star
during this low-pulse-fraction period.

As can be seen from the zoomed plot, all three events started about 
10 days after the periastron, a feature already noted by \citet{Aba20}.
Similar starting time was also true for the two spin-up events reported by \citet{Koh97}. 
The meaning of this start time will be discussed in \S. 4.

\section{"Pre"-periastron flare of GX 301-2}

Early observations of GX 301-2 revealed that the X-ray fluxes of GX 301-2 show an orbital 
flare about 1.4 days before the periastron passage \citep[e.g.][]{Sat86}, a very 
particular feature among X-ray pulsars. Many continuous monitoring of GX 301-2 in different energy 
bands are available in recent years, such as Swift/BAT and MAXI, and they allow more detailed studies
of the behavior of the pre-periastron flare of GX 301-2. 
In Figure 3, we plot the orbital profile of the fluxes of Swift/BAT 15-50 keV, MAXI 2-10 keV 
and 10-20 keV, and of the pulsed fluxes within 12-50 keV estimated by Fermi/GBM.
The MAXI data are taken
from its Gas Slit Camera\footnote{http://maxi.riken.jp/}.
We use the orbital parameter of period $P=41.472$ days with
$\dot{P}=-3.7\times10^{-6}$, 
and a periastron time $T_{per}=53531.65$ (MJD) given by \citet{Dor10}.
The no spin-up period between MJD 55400 and 58400 is adopted.
All the profiles are normalized to the Swift/BAT profile at phase 0.5, and 100 orbital 
phase bins are used.

\begin{figure}
	\hspace{-1.0cm}
	\includegraphics[width=3.7in]{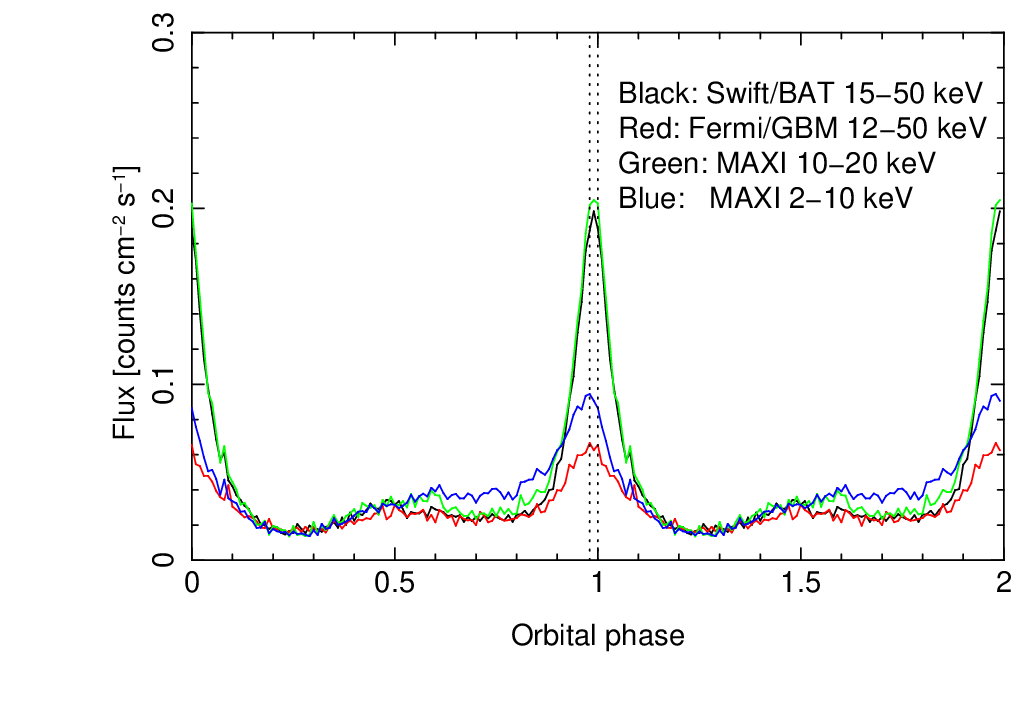}
\caption{Orbital profiles of the fluxes of Swift/BAT 15-50 keV (black), MAXI 10-20 keV (green),
MAXI 2-10 keV (blue), and the pulsed fluxes of Fermi/GBM within 12-50 keV (red).
Two dotted lines around phases of 0.98 and 1 are plotted to guide eyes.
}
\end{figure}

It can be seen that the Swift/BAT 15-50 keV and MAXI 10-20 keV profiles are quite similar,
both of which show a prominent peak around phase 0.99, only 0.4 days before the periastron passage.
The peak fluxes are about 10 times the lowest values around phases between 0.2 and 0.3.
The MAXI 2-10 keV profile is peaked around phase 0.98, but it is skewed to earlier phases, with a 
centroid around phase 0.97. The peak 2-10 keV fluxes are about 4.5 times 
the lowest values. In contrast, the peak pulsed fluxes within 12-50 keV are only 3 times 
the lowest values,
indicating that the pulsed fractions around the flux peak are only 30\% of those around the 
low flux periods.
These results show that the intrinsic high-energy X-ray emission (15-50 keV) of 
GX 301-2 peaked around orbital phase 0.99, very close to the periastron passage, 
and the "pre"-periastron flare is only significant for lower energies (2-10 keV), which are
more easily absorbed by matter around GX 301-2.

The drop of pulse fraction near periastron has already been shown by \citet{Nab19}, and 
they noted that it may be caused by scattering in the dense wind environment.
To further illustrate the nature of the drop of pulsed fraction near periastron, 
we extract the averaged Swift/BAT spectra for orbital phases within 0.1-0.9 and 0.9-1.1, respectively, 
using the snapshot light curves of the eight bands of BAT. The spectra are presented 
in Figure 4, and the spectrum within phase 0.1-0.9 is normalized to that of phase 0.9-1.1
at 14-20 keV. It can be seen that the spectrum near periastron is steeper than that away from 
periastron. That is, if the intrinsic X-ray emission is similar for all phases, the higher 
energy photons near periastron are more scattered to lower energies. This is the typical 
behavior of Compton scattering process. Indeed, Compton shoulder of the Fe \Ka line has been reported 
for GX 301-2 with \cha observations near periastron \citep{Wat03,Liu18}. Considering the absorption 
column density of GX 301-2 near periastron is generally larger than $10^{24}$\,cm$^{-2}$, 
the Compton scattering optical depth is larger than 1, a large part of the intrinsic high 
energy emission will be scattered, and the observed high energy X-ray photons will also 
contain a large amount of scattered photons. 
The surface of the optical star could also be an origin of the scattering medium.
Therefore, Compton scattering will reduce 
the pulse fraction near periastron, just as observed.
Compton reflection might also affect 
the measurement of the spin derivatives near periastron, which will be discussed in next section.

\begin{figure}
	\hspace{-1.0cm}
	\includegraphics[width=3.7in]{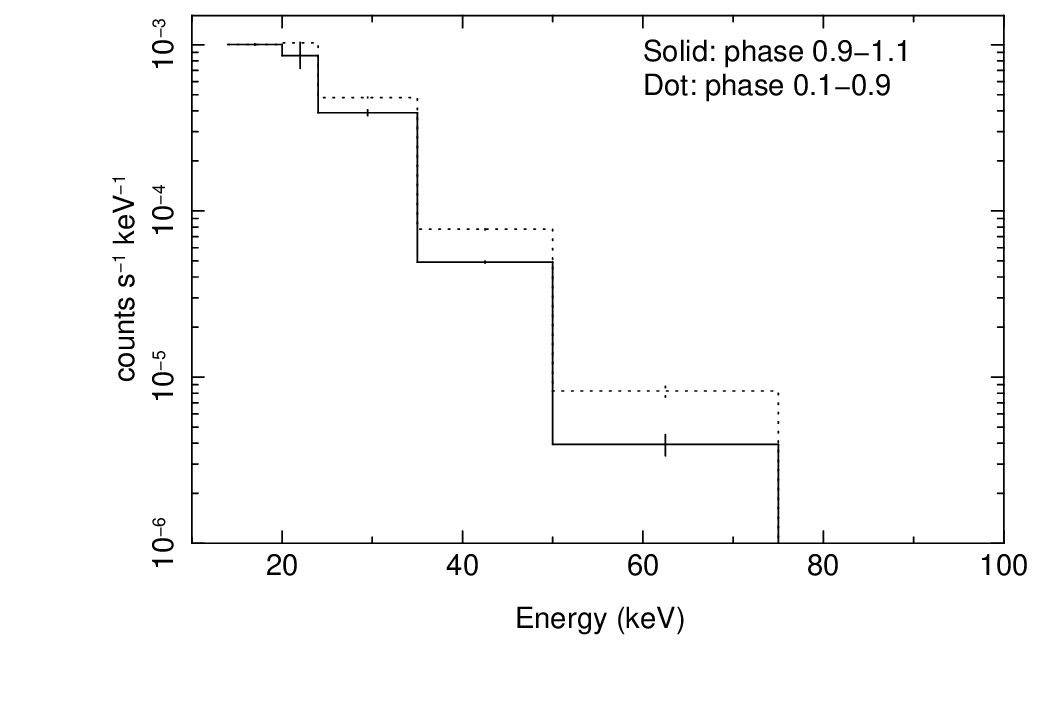}
\caption{Swift/BAT spectra of GX 301-2 extracted from phases within 0.9-1.1 (near periastron)
	and with 0.1-0.9. The steeper spectral shape of the spectrum within phase 0.9-1.1 indicates
	Compton scattering of the intrinsic high energy X-ray emission near periastron.
				}
\end{figure}

\section{Conclusion and Discussion}

We studied the three rapid spin-up events of GX 301-2 detected by Fermi/GBM in last ten years
together with Swift/BAT and MAXI data. The spin derivatives of GX 301-2 during the no spin-up 
period show no correlation with the Swift/BAT 15-50 keV fluxes, consistent with a wind accretion
mode; while the spin derivatives during the rapid spin-up periods
are correlated with the Swift/BAT fluxes, following a relation of 
$\dot{\nu}\propto F^{0.75\pm0.05}$. Such a correlation shows that the accreted matter 
has the same flow direction as the rotation of the neutron star, consistent with accretion 
of a prograde disk to the neutron star \citep{PR72,RJ77}. 
These results confirm the previous idea that the rare spin-up events of GX 301-2 are due to a transient
disk \citep{Koh97}.

All the rapid spin-up events of GX 301-2 started about 10 days after the periastron passage.
We note that such a timescale is corresponding to the time needed for the tidally stripped gas
to reach the neutron star. As shown in hydrodynamic simulations by \citet{Lay98}, 
the tidally stripped gas 
reaches the neutron star around phase 0.25, just about 10 days after the periastron passage.
The optical star Wray 977 has a radius
of $62 R_\odot$, while the Roche-Lobe radius near periastron is about $66 R_\odot$
estimated with the formula given by \citet{Sep07}.
As a result, Wray 977 is underfill its Roche-Lobe radius, consistent with the fact 
that GX 301-2 is dominated by 
wind accretion. Nevertheless, the radius of Wray 977 is close enough to the Roche-Lobe radius
near periastron, making it possible to transport some tidally stripped gas to the 
neutron star occasionally, to produce the observed spin-up events. 
Such a transport process might be caused by the instability of Wray 977 or other unknown effects.

Optical spectral observations of Wray 977 favor a slow rotation of Wray 977, with a period 
comparable to the orbital period of GX 301-2 \citep{Par80, Kap06}. Therefore, Wray 977 is 
rotating much slower ($\sim1/3$) 
compared to the angular velocity of the neutron star near periastron. 
The simulation by \citet{Lay98} predicted a spin-down event following the spin-up event, 
which is not observed for the spin-up events (but \citet{Mon20} discovered an orbital 
spin-up to spin-down reversal near periastron). The predicted spin-down is due to their 
assumption of synchronous
rotation of the optical star near periastron, which makes the stripped gas rotating faster 
than the angular velocity of the neutron star after periastron and leads to a reversed 
accretion direction after the stripped stream passes over the neutron star. 
The slow rotation of Wray 977 makes the passage of stripped gas stream over the 
neutron star unlikely, consistent with the fact that only spin-up events are observed. 

For a scenario of accretion of tidally stripped gas to the neutron star, 
the stripped matter generally has angular momentum in the direction of the orbital motion, 
as in a Roche-Lobe-like overflow.
The stripped matter might pass over the neutron star only when it obtains a high radial 
velocity and when the optical star rotates faster than the angular velocity of the neutron 
star near periastron. The slow rotation nature of Wray 977 makes a retrograde disk 
unlikely. Therefore, the observed spin-up events and the slow rotation of GX 301-2 favor
a prograde neutron star of GX 301-2.
This is in contrast to the idea of a retrograde neutron star proposed
by \citet{Mon20} to explain the orbital spin-up to spin-down reversal near periastron. 
Their idea is based on the stream model invoked to explain the pre-periastron flare
of GX 301-2 \citep{LK08}.
We found that the flare of the intrinsic X-ray emission (high energy) of GX 301-2 happened 
around phase 0.99, only 0.4 days before the periastron, while the flare of low energies (2-10 keV) 
happened about 1.4 days before the periastron. 
The preceding low-energy flare can 
be explained as due to stronger absorption of the intrinsic X-ray emission closer to the periastron.
This finding weakened the need of the stream model for GX 301-2. In addition, the stream model 
requires that the optical star rotates twice faster as the orbital motion, in conflict with the slow 
rotation of Wray 977. 

The pulse fraction of GX 301-2 near periastron is reduced heavily (30\%) compared with 
other orbital phases. Such a drop of pulse fraction can be explained by Compton scattering 
from the dense material around Wray 977 and/or from the surface of Wray 977. Compton 
process may also provide a clue to the orbital spin reversal near periastron
discovered by \citet{Mon20}. The periastron distance between the neutron star and the surface 
of Wray 977 is only about $30 R_\odot$ (70\,lt-s), while around phase 0.9 the distance is about
$80 R_\odot$ (185\,lt-s). These distances are smaller than the light travel distances
within the pulse period of GX 301-2, 680\,s. Therefore, Compton reflected photons
from the optical companion might affect the 
observed pulse profile heavily and affect the measurement of spin derivatives
near periastron. Detailed modelling of reprocession and transport of hard X-ray emission is needed 
to test whether Compton reflection is responsible for the observed orbital spin
reversal of GX 301-2.


\section*{Acknowledgements}
I thank the referee for helpful comments, Jenke Peter A. for help on Fermi/GBM data,
Lien Amy Yarleen for help on 
Swift/BAT data, Zheng Xueying, Liao Zhenxuan, Lu Youjun, Li Xiangdong, and 
Epili Prahlad for helpful discussions. 
This research is supported by National Natural Science Foundation of China (11773035 and U1938113)
This research used data obtained with the Fermi/GBM, Swift/BAT, and MAXI.

\section*{Data availability}
The data underlying this article are publicaly available from the web as listed in the footnote.
\bibliographystyle{mn2e}

\appendix

\end{document}